\documentclass[english,prl,aps,twocolumn]{revtex4-2}
\usepackage{amsmath}
\usepackage{amssymb}
\usepackage{graphicx}
\usepackage{babel}
\usepackage{array}
\usepackage{verbatim}
\usepackage{xcolor}
\usepackage[colorlinks=true, pdfstartview=FitV, linkcolor=blue, citecolor=blue, urlcolor=blue]{hyperref} 

\def\CrO      {{Cr\textsubscript{2}O\textsubscript{3}}}

\newcommand{\ee}[1]{\cdot10^{#1}}
\newcommand{\mr}[1]{\mathrm{#1}}
\newcommand{\unit}[1]{\,\mathrm{#1}}
\newcommand{\um}{\,\mu{\rm m}}
\newcommand{\us}{\,\mu{\rm s}}
\newcommand{\uT}{\,\mu{\rm T}}
\newcommand{\uA}{\,\mu{\rm A}}
\newcommand{\uW}{\,\mu{\rm W}}

\newcommand{\muB}{\mu_{\rm B}}
\newcommand{\rtHz}{\sqrt{\mr{Hz}}}

\newcommand{\ye}{\gamma_\mr{e}}

\newcommand{\Bext}{B_\mr{ext}}

\newcommand{\vecB}{B}

\newcommand{\fc}{f_\mr{TF}}
\newcommand{\ko}{k_0}
\newcommand{\kx}{k_x}
\newcommand{\ky}{k_y}
\newcommand{\kr}{k_r}
\newcommand{\ta}{\tau}

\newcommand{\Ttwo}{T_{2}}
\newcommand{\xo}{x_0}
\newcommand{\xosc}{x_\mr{osc}}

\newcommand{\phim}{\phi^{(\text{measured})}}
\newcommand{\xbi}{\chi_\mr{Bi}}
\newcommand{\xpd}{\chi_\mr{Pd}}
\newcommand{\BT}{B_\mr{T}}
\newcommand{\hatBT}{\hat{B}_\mr{T}}

\begin{document}
	\title{Scanning gradiometry with a single spin quantum magnetometer}
		
	\author{W.~S.~Huxter$^{1,\dagger}$, M.~L.~Palm$^{1,\dagger}$, M.~L.~Davis$^{1,\dagger}$, P.~Welter$^{1}$, C.-H.~Lambert$^{2}$, M.~Trassin$^{3}$, and C.~L.~Degen$^{1,4}$}
  
	\affiliation{$^1$Department of Physics, ETH Zurich, Otto Stern Weg 1, 8093 Zurich, Switzerland.}
	\affiliation{$^2$Department of Materials, ETH Zurich, H\"{o}nggerbergring 64, 8093 Zurich, Switzerland.}
	\affiliation{$^3$Department of Materials, ETH Zurich, Vladimir Prelog Weg 1-5/10, 8093 Zurich, Switzerland.}
  \affiliation{$^4$Quantum Center, ETH Zurich, 8093 Zurich, Switzerland.}

  \email{degenc@ethz.ch}
  \thanks{$^\dagger$These authors contributed equally.}
	
	\begin{abstract}
	Quantum sensors based on spin defects in diamond have recently enabled detailed imaging of nanoscale magnetic patterns, such as chiral spin textures, two-dimensional ferromagnets, or superconducting vortices, based on a measurement of the static magnetic stray field. 
	Here, we demonstrate a gradiometry technique that significantly enhances the measurement sensitivity of such static fields, leading to new opportunities in the imaging of weakly magnetic systems.  Our method relies on the mechanical oscillation of a single nitrogen-vacancy center at the tip of a scanning diamond probe, which up-converts the local spatial gradients into ac magnetic fields enabling the use of sensitive ac quantum protocols.  We show that gradiometry provides important advantages over static field imaging: (i) an order-of-magnitude better sensitivity, (ii) a more localized and sharper image, and 	(iii) a strong suppression of field drifts.  We demonstrate the capabilities of gradiometry by imaging the nanotesla fields appearing above topographic defects and atomic steps in an antiferromagnet, direct currents in a graphene device, and para- and diamagnetic metals.  
	\end{abstract}
	
	\date{\today}
	\maketitle

Nanoscale magnetic stray fields appearing at surfaces and interfaces of magnetically-ordered materials provide important insight into the local spin structure.  Such stray fields are present, for example, above magnetic domains and domain walls \cite{rondin12}, magnetic vortices \cite{shinjo00}, spin spirals \cite{cox63}, skyrmions \cite{yu10}, or topographic steps and defects \cite{ravlic03}, and often accompany other types of ordering, like ferroelectricity \cite{fiebig02}.  Similar stray fields also appear near flowing currents \cite{chang17,tetienne17} or materials with a magnetic susceptibility \cite{bluhm09spin}.  Therefore, stray field measurements are a general and versatile tool to study local material or device properties.

While the magnetic imaging of ferromagnetic and ferrimagnetic textures is well-established \cite{hopster03,freeman01,hartmann99}, detection of the much weaker stray fields of, for example, antiferromagnets, multiferroics or nanoscale current distribution is a relatively new development.  Quantum magnetometers based on single nitrogen-vacancy (NV) centers have recently led to exciting advances in this direction \cite{gross17,appel19,wornle19,wornle21,hedrich21,jenkins20,ku20}.  In their standard configuration, NV magnetometers image stray fields by scanning a sharp diamond probe over the sample surface and monitoring the static shift of the NV spin resonance frequency \cite{rondin14,degen08apl}.  State-of-the-art scanning NV magnetometers reach a sensitivity to static fields of a few $\uT/\rtHz$ \cite{wornle21,sun21}.  This sensitivity is sufficient for imaging the domain structure of monolayer ferromagnets \cite{thiel19,sun21,fabre21} and uncompensated antiferromagnets \cite{appel19,wornle19,wornle21,hedrich21}, however, it remains challenging to detect the even weaker stray fields of isolated magnetic defects, spin chains or ideally compensated antiferromagnets.  While higher sensitivities, on the order of $50\unit{nT/\rtHz}$, have been demonstrated using dynamic (ac) detection of fields \cite{vool21,palm22}, this approach is limited to the few systems whose magnetization can be modulated, like isolated spins \cite{rugar04,grinolds14}.

Here, we demonstrate a gradiometry technique for highly sensitive imaging of static magnetization patterns.  Our method relies on up-conversion of the local spatial gradient into a time-varying magnetic field using mechanical oscillation of the sensor combined with sensitive ac quantum detection.  This operating principle is well known from dynamic force microscopy \cite{hillenbrand00,sahin07} and has been explored with NV centers \cite{degen08apl,hong12,kolkowitz12science,teissier14,esquinazi15,wood18} and other magnetic sensors \cite{uri20nphys,wyss21} in various forms, however, it has not been realized for imaging general two-dimensional magnetic samples.  As a demonstration, we show that scanning gradiometry is able to resolve the nanotesla magnetic stray fields appearing above single atomic terraces in antiferromagnetic \CrO. Imaging of nanoscale current patterns, magnetic susceptibility in metals, and reconstruction of field maps from gradiometry data are also demonstrated.

\vspace{0.5cm}
\textit{Gradiometry technique -- }
%
\begin{figure}[t]
	\includegraphics[width=0.48\textwidth]{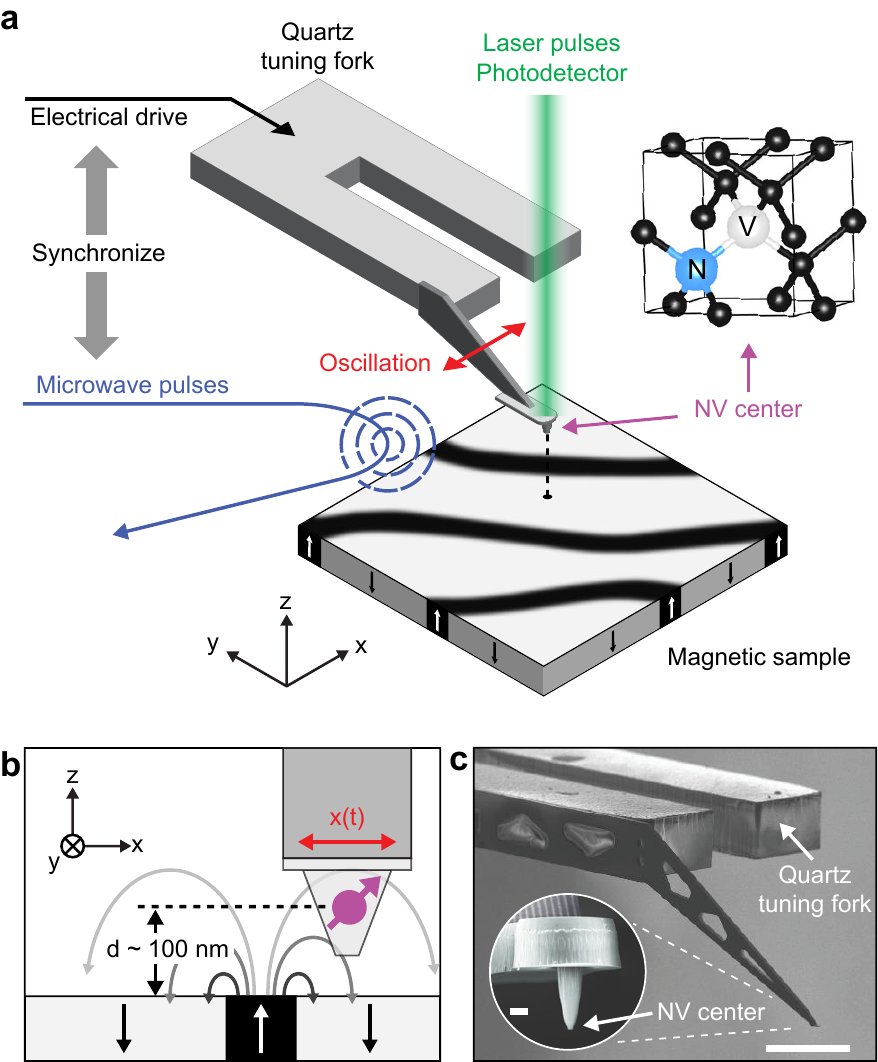}
	\caption{{\bf Experimental set-up}.
	{\bf a}, Schematic of the scanning gradiometer. A single NV center located at the apex of a diamond tip is oscillated in shear-mode using a quartz tuning fork.  The microwave pulse generation is synchronized, via a lock-in controller, with the electrical drive of the tuning fork.  An objective located above the NV (not shown) is used to apply laser pulses and collect the NV photo-luminescence.  A three-axis piezo stage (not shown) is used to position and scan the sample surface under the sensor.	
	{\bf b}, Detail showing the orientations of sample, tip and direction of oscillation.
	{\bf c}, Scanning electron micrograph of a quartz tuning fork and diamond tip (inset).  Scale bars, $200\unit{\um}$ and $1\unit{\um}$ (inset).
	\label{fig1}
	}
\end{figure}
%
\begin{figure}[t]
	\includegraphics[width=0.48\textwidth]{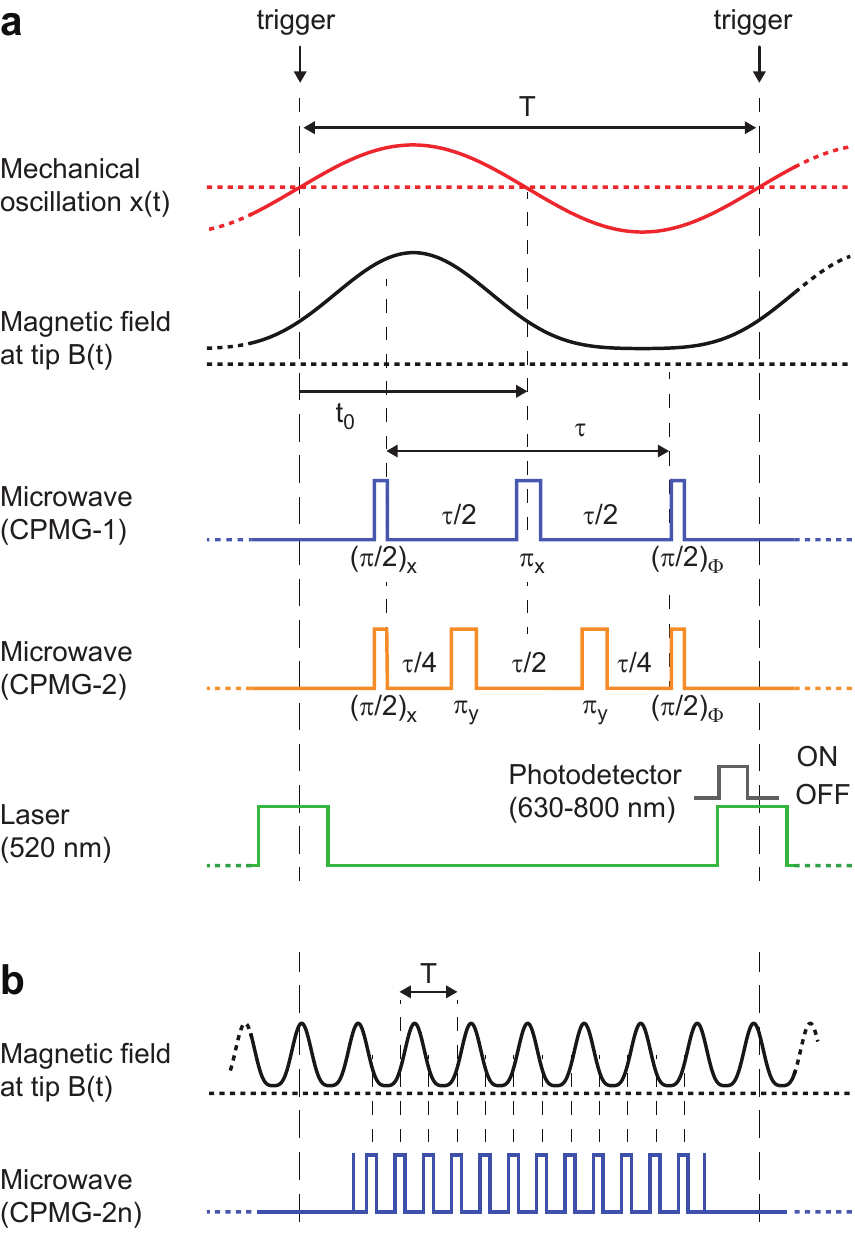}
	\caption{{\bf AC quantum sensing protocol for gradient detection}.
	{\bf a}, Timing diagram of sensor oscillation (red), resulting magnetic field at the sensor location (black), microwave pulse sequences for first and second harmonic (blue and orange), laser pulses (green) and photodetector collection window (gray).
	Pulse indicators such as $(\pi)_\mr{x}$ denote (angle)$_\text{axis}$ of the spin rotation.  $T=1/\fc$ is the oscillation period, $\tau < T$ is the sensor interaction time, and $t_0$ the trigger delay.  We use a tuning fork with $\fc \sim  32\unit{kHz}$.  Durations of pulses are $\sim 2\unit{\us}$ (laser) and $\sim 100\unit{ns}$ (microwave $\pi$-rotation).
	{\bf b}, Scheme for detecting the first harmonic signal over $n$ oscillation periods in the regime $\tau\gg T$. Here, $\tau = nT$ and the number of $\pi$-pulses is $2n$.
	\label{fig2}
	}
\end{figure}
The operating principle of our gradiometry technique is shown in Fig.~\ref{fig1}a.  Our scanning magnetometer set-up consists of a sample plate that is scanned underneath a diamond probe containing a single NV center at the tip apex formed by ion implantation \cite{qzabre}.  The diamond tip is attached to the prong of a quartz tuning fork (Fig.~\ref{fig1}) providing atomic force microscopy (AFM) position feedback \cite{giessibl98}.  The microscope apparatus additionally includes an objective to optically polarize and read-out the NV center spin state and a bond wire acting as a microwave antenna for spin manipulation (see Methods for details).  In the conventional mode of operation, we record a spin resonance spectrum at each pixel location to build up a map of the sample's static stray field \cite{wornle21}.

To implement the gradiometer, we mechanically oscillate the NV in a plane parallel to the sample (shear-mode) by electrically driving the tuning fork at a fixed amplitude $\xosc \sim 10-70\unit{nm}$.  The NV center now experiences a time-dependent field given by
\begin{align}
\vecB(x(t))
  &= \vecB(\xo) \nonumber \\
	&+ \left.\frac{\partial \vecB}{\partial x}\right|_{x=\xo} \xosc \sin(2\pi\fc t) \nonumber \\
	&+ \left.\frac{\partial^2 \vecB}{\partial x^2}\right|_{x=\xo} \frac{\xosc^2}{2} \sin^2(2\pi\fc t)
	+ \dots
\end{align}
where $\vecB$ is the vector component of the sample's stray field along the NV anisotropy axis (Methods), and where $x(t) = \xo + \xosc\sin(2\pi\fc t)$ describes the mechanical oscillation around the center location $\xo$ with frequency $\fc$.  The amplitudes of the $0\fc$, $1\fc$ and $2\fc$ harmonics in leading orders of $\xosc$ are given by
\begin{subequations}
	\label{eq:taylor}
	\begin{align}
		B_0 &= B(\xo)  \label{eq:b0} \\
		B_1 &= \xosc  \left.\frac{\partial B}{\partial x}\right|_{x=\xo} \label{eq:b1} \\
		B_2 &= \frac12 \xosc^2 \left.\frac{\partial^2 B}{\partial x^2}\right|_{x=\xo} \label{eq:b2} 
	\end{align}
\end{subequations}
and are therefore proportional to the static field, gradient, and second derivative, respectively.  The series expansion in Eq.~(\ref{eq:taylor}) is accurate for oscillation amplitudes $\xosc$ smaller than the scan height, typically $d\sim 100\unit{nm}$.

To detect the harmonics of $B(t)$, we synchronize the mechanical oscillation with a suitable ac quantum sensing sequence~\cite{taylor08,delange11}, shown in Fig.~\ref{fig2}a.  Quantum sensing sequences measure the quantum phase accumulated by the coherent precession of a superposition of spin states during an interaction time $\tau$ \cite{degen17}.  To measure the $n\fc$ harmonic, we invert the spin precession $n$ times during one mechanical oscillation period using microwave $\pi$-pulses (Carr-Purcell-Meiboom-Gill (CPMG-$n$) sequence \cite{meiboom58,degen17}).  The pulse protocols for the first (CPMG-1) and second (CPMG-2) signal harmonic are shown in Fig.~\ref{fig2}a.  The quantum phase accumulated for the first harmonic is given by 
\begin{align}
\phi
  &= \int\displaylimits_{t=t_0-\tau/2}^{t_0} \ye B(t) \mr{d}t
	 -\int\displaylimits_{t=t_0}^{t_0+\tau/2} \ye B(t) \mr{d}t \nonumber \\
  &= \ye B_1 \, \frac{2\sin^2(\pi\fc\ta/2)}{\pi\fc}
\label{eq:phi}
\end{align}
where $\ye = 2\pi\times 28\unit{GHz/T}$ is the gyromagnetic ratio of the NV electronic spin and $t_0 = T/2$ (see Fig.~\ref{fig2}a).  The derivation of Eq.~(\ref{eq:phi}) and general expressions for higher harmonics and for sequences that are off-centered with respect to the tuning fork oscillation ($t_0 \neq T/2$) are given in Supplementary Note 1.
To determine $\phi$ experimentally, we measure the photo-luminescence (PL) intensity $C_\Phi$ as a function of the phase $\Phi = x, y, -x, -y$ of the last microwave $\pi/2$ pulse (Fig.~\ref{fig2}a).  From the four PL signals $C_\Phi$ we then extract the phase using the two-argument arc tangent (see Methods),
\begin{align}
\phim = \text{arctan}\left(\frac{C_{-y}-C_{y}}{C_x-C_{-x}}\right) \ .
\label{eq:phim}
\end{align}
This four-phase readout technique has the advantage that the phase can be retrieved over the full $2\pi$ range with uniform sensitivity \cite{ku20,palm22}.  From $\phim$ and Eq.~(\ref{eq:phi}) we then compute the gradient field $B_1$.  Since we can only measure $\phi$ modulo $2\pi$, a phase unwrapping step is necessary for large signals that exceed the range $[-\pi;\pi)$ \cite{palm22}.  

For NV centers with long coherence times ($\Ttwo\gg T$), the sensitivity can be improved further by accumulating phase over multiple oscillation periods using CPMG-$2n$ sequences (Fig.~\ref{fig2}b).  The simplest way to meet the $\Ttwo\gg T$ condition is to use a mechanical oscillator with higher resonance frequency.  Alternatively, as demonstrated in this work, a higher mechanical mode of the tuning fork can be employed \cite{zeltzer07}.  High frequency detection has the added advantage that the interval between $\pi$-pulses (given by $1/(2\fc)$) becomes very short, which makes dynamical decoupling more efficient and in turn leads to improved sensitivity \cite{delange11,ryan10}.  We demonstrate single and multi-period detection schemes with sensitivities of $\sim 120\unit{nT/\rtHz}$ and $\sim 100\unit{nT/\rtHz}$, respectively, in Supplementary Notes 2 and 3.

\vspace{0.5cm}
\textit{Demonstration of scanning gradiometry -- }
%
\begin{figure*}
	\includegraphics[width=1.00\textwidth]{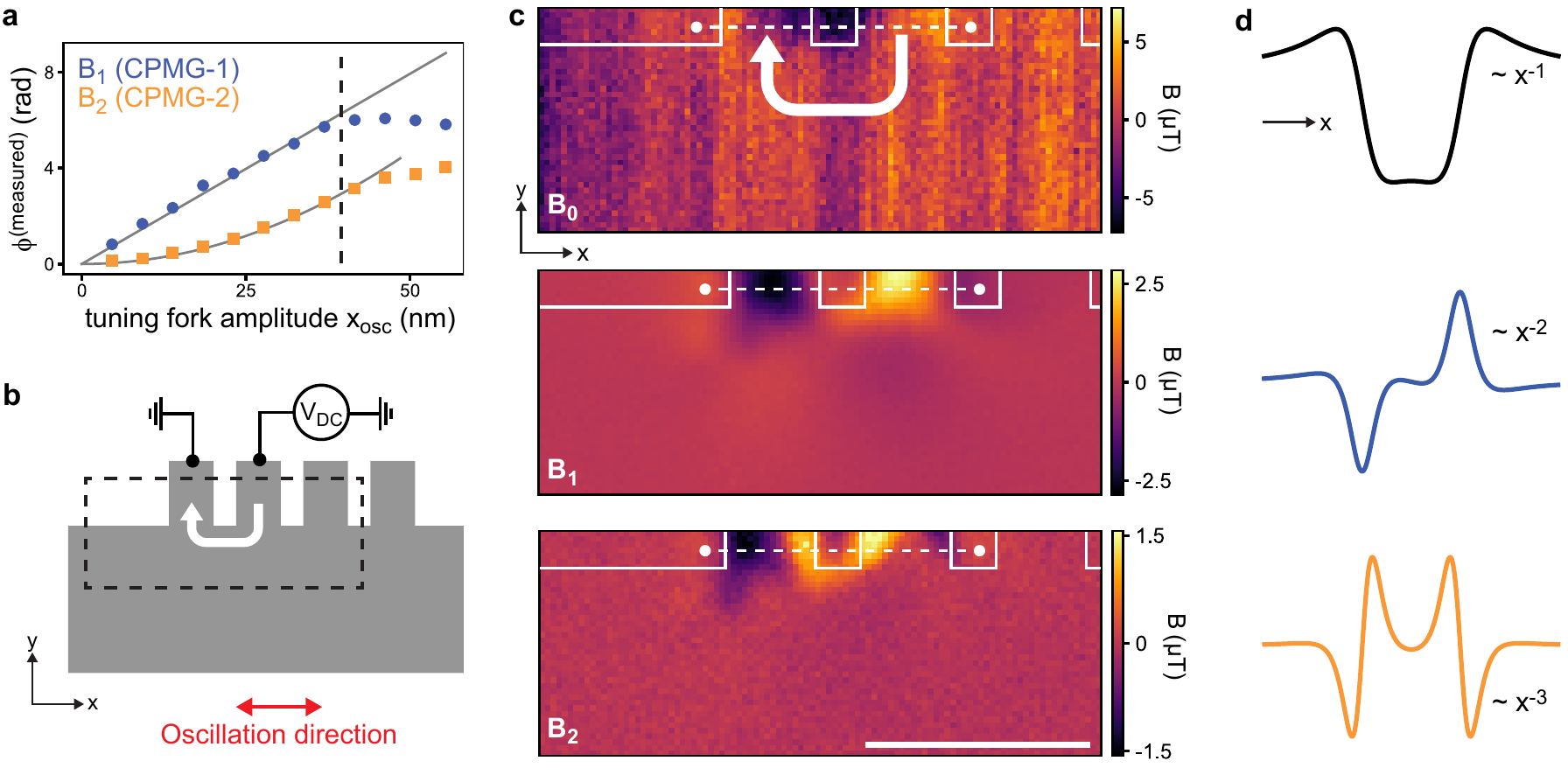}
	\caption{{\bf Two-dimensional scanning gradiometry of current flow in a graphene test geometry.}
		{\bf a}, Measured and unwrapped quantum phase $\phim$ plotted as a function of tip oscillation amplitude $\xosc$ for the first (blue) and second (orange) harmonic.  Solid lines are $\propto\xosc$ and $\propto\xosc^2$, respectively.  Dashed line indicates maximum oscillation amplitude before the Taylor approximation breaks down due to the onset of spatial averaging.
		{\bf b}, Schematic of the bilayer graphene device. A direct current of $I_\mr{dc}\approx 5.3\unit{\uA}$ (white arrow) is applied between the indicated contacts (width $\sim 400 \unit{nm}$).
		{\bf c}, Static field map $B_0$, gradient map $B_1$, second derivative map $B_2$ from the area indicated in {\bf b}.  Contours reflect the device layout. Dwell time is $30\unit{s}$ per pixel.  Scale bar, $1 \unit{\um}$. 
		{\bf d}, Analytical spatial profiles for the $z$ component of the magnetic field, gradient and second derivative generated across the two contacts (dotted lines in {\bf c}).  The far-field decay with $x$ is also indicated.
		\label{fig3}
  }
\end{figure*}
We begin measurements by calibrating the oscillation amplitude $\xosc$ and verifying Eqs.~(\ref{eq:taylor},\ref{eq:phi}).  We characterize the gradiometry phase detection by parking the tip over a fixed position on a magnetic test sample and measuring the $B_1$ and $B_2$ signals as a function of the oscillation amplitude $\xosc$ (see Methods and Supplementary Note 4 for characterization and calibration details).  Fig.~\ref{fig3}a confirms that the signals grow as $B_1 \propto \xosc$ and $B_2 \propto \xosc^2$, as expected from the Taylor expansion (Eq.~\ref{eq:taylor}).  Further, to avoid mixing of mechanical and field harmonic signals through surface interactions, we retract the tip by approx. $20\unit{nm}$ from the contact point while measuring (Supplementary Note 5).

We establish two-dimensional imaging by detecting the stray field from a direct current flowing in a bilayer graphene device (Fig.~\ref{fig3}b).  This device provides an ideal test geometry because the magnetic stray field and gradient can be directly compared to the analytical model.  Fig.~\ref{fig3}c presents gradiometry maps $B_1$ and $B_2$ and, for comparison, a static field map $B_0$ recorded using a standard dc technique \cite{dreau11} with no tuning fork oscillation.  Note that all images are recorded with the same imaging time and pixel size.
The figure immediately highlights several advantages of gradiometry versus static field imaging:
First, the signal-to-noise ratio (SNR) is strongly enhanced due to the more sensitive ac detection, despite of a lower absolute signal.  Second, magnetic drift throughout the imaging time (several hours) is present in the static image but suppressed in the gradiometry images.  Third, because gradient fields decay quickly with distance ($B_1\propto x^{-2}$ and $B_2\propto x^{-3}$ compared to $B_0 \propto x^{-1}$, see Fig.~\ref{fig3}d and Supplementary Note 6), they offer a higher apparent image resolution and are thus easier to interpret.

\vspace{0.5cm}
\textit{Imaging of antiferromagnetic surface texture -- }
%
%
\begin{figure*}
	\includegraphics[width=1.0\textwidth]{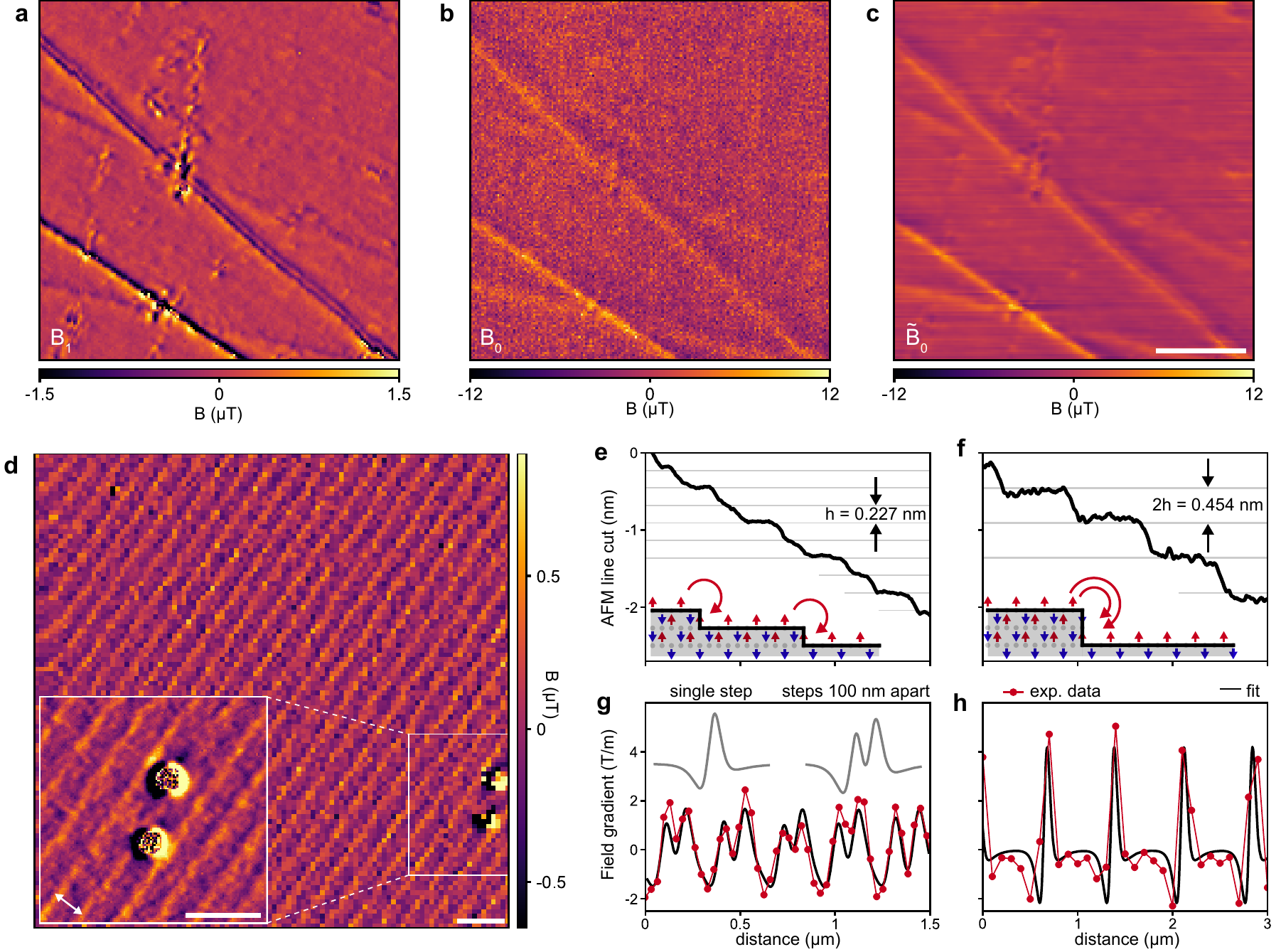}
	\caption{{\bf Topographic fields from antiferromagnetic \CrO.}
		{\bf a}, Gradiometry map $B_1$ of topographic defects from a polished \CrO\ single crystal.  Dwell time is $10\unit{s}$ per pixel and $\xosc=23\unit{nm}$.
		{\bf b}, Corresponding static field map $B_0$.  Dwell time is $1\unit{s}$ per pixel.
		{\bf c}, Improved static field map $\tilde B_0$ obtained from a weighted average of {\bf a} and {\bf b} (see Methods).  The noise standard deviation is reduced from $1.8\unit{\uT}$ ({\bf b}) to $90\unit{nT}$ in the oscillation direction ($x$-direction) and $340\unit{nT}$ in the $y$-direction ({\bf c}).  The directional sensitivity could be avoided using tapping mode oscillation (see Methods).  When normalized by measurement time, $B_1$ is more sensitive than $B_0$ by a factor larger than 8. 
		{\bf d}, $B_1$ image of the stray field generated by atomic steps on an as-grown \CrO(0001) surface.  Two magnetic defects which produce much larger gradient signals are also visible.  Inset is a detail of defects and atomic steps recorded with higher pixel resolution.  Distance between steps is ca. $400 \unit{nm}$ (arrow).
		Scale bars in {\bf a-d}, $1 \unit{\um}$.
		{\bf e}-{\bf f}, AFM line scans revealing mono- and diatomic steps with different terrace sizes. $h$ is the distance between neighboring O-planes along the \CrO\ $c$-axis \cite{fiebig96thesis,he10}. Insets show sketches of atomic steps and up-down ordering of Cr$^{3+}$ moments (red, blue).  Curved arrows are field lines.
		{\bf g}-{\bf h}, Gradiometry line scans acquired over different areas on the as-grown \CrO\ surface. The profiles are fitted to a stray field model involving atomic steps (solid lines).
		$\xosc=46\unit{nm}$ in \textbf{d}, \textbf{g}, and \textbf{h}.
		\label{fig4}
  }
\end{figure*}
We now turn our attention to the imaging of stray fields above antiferromagnetic materials, focusing on the archetypical model system \CrO.  Antiferromagnets represent a general class of weakly magnetic materials that are both challenging to image by existing techniques \cite{fiebig95,scholl00,wu11} and of key importance for understanding multiferroicity and topological magnetism in the context of antiferromagnetic spintronics \cite{cheong20}.  Although antiferromagnets are nominally non-magnetic, weak stray fields can appear due to uncompensated moments at domain walls \cite{appel19,wornle19,wornle21}, spin spirals \cite{gross17}, topographic steps \cite{hedrich21}, or surface roughness and defects.  The latter are particularly difficult to detect, yet play an important role in the pinning of domain walls \cite{hedrich21} and establishing an exchange bias at material interfaces \cite{nogues99}.  Careful study of these weaker fields is therefore an important avenue for quantifying the interplay between surface roughness, stray field structure, local magnetization and domain wall behavior.

In Fig.~\ref{fig4}a we show a gradiometry image of a polished \CrO\ single-crystal \cite{wornle21}.
\CrO\ has a layered structure of out-of-plane Cr$^{3+}$ moments at the (0001) surface \cite{fiebig96thesis,he10} that lead to stray fields at topographic steps \cite{wornle21,hedrich21}.  These stray fields are proportional to the step height.  Indeed, Fig.~\ref{fig4}a reveals a rich variety of magnetic anomalies on this surface, including two $\sim 5\unit{nm}$ deep and $\sim 200\unit{nm}$ wide topographic trenches introduced by polishing, a number of point defects, and general texture of the $\sim$2\,nm-rms surface roughness (see Fig.~S11 for topographic characterization).  A quantitative calculation of the expected stray field maxima shows that the magnetic anomalies are well explained by the surface topography (Supplementary Note 7).  Except for the trenches, these surface defects are not visible in the static field image (Fig.~\ref{fig4}b).  Note that it is possible to convert the gradiometry map into an improved static field map through integration and weighted averaging in $k$-space (Fig.~\ref{fig4}c and Methods).

Next, we show that gradiometry is sufficiently sensitive to detect single atomic surface steps of the layered \CrO(0001) surface.  Fig.~\ref{fig4}d displays a gradiometry image from a second \CrO\ single crystal with an as-grown (unpolished) surface \cite{fiebig96thesis}.  We observe a striking pattern of regular stripes separated by a few hundred nanometers and an approximate amplitude of $B_1 \sim 250\unit{nT}$.  The pattern extends over the entire crystal surface with different stripe separations and directions in different regions of the sample (Fig.~S12).  By converting $B_1$ measurements into local gradients we deduce height changes $<1\unit{nm}$ (Supplementary Note 7), suggesting that the repeating striped patterns are caused by single ($h = 0.227 \unit{nm}$), or multiple atomic step edges. To corroborate, we correlate the magnetic features to the AFM topography of the sample (Fig.~S12).
Figs.~\ref{fig4}e and \ref{fig4}f show line cuts perpendicular to the stripe direction in two regions of the sample surface, and reveal the presence of both mono- and diatomic step edges.  The gradiometry data (Fig.~\ref{fig4}g,h), are well fitted by a simple model (see Methods), producing a fitted surface magnetization of $\sigma_z = 2.1 \pm 0.5 \unit{\mu_B/nm^2}$, in good agreement with earlier data \cite{appel19,wornle21}.  Together, our findings unambiguously confirm the stepped growth of the \CrO(0001) surface \cite{he10}.  To our knowledge, our work reports the first magnetic stray field imaging of atomic steps on an antiferromagnet.
It opens a complementary path to existing atomic-scale techniques, such as spin-polarized scanning tunneling microscopy and magnetic exchange force microscopy \cite{wiesendanger09}, without sharing the needs for an ultra-high vacuum, a conducting surface, or cryogenic operation.

\vspace{0.5cm}
\textit{Imaging of magnetic susceptibility -- }
%
\begin{figure}
	\includegraphics[width=0.50\textwidth]{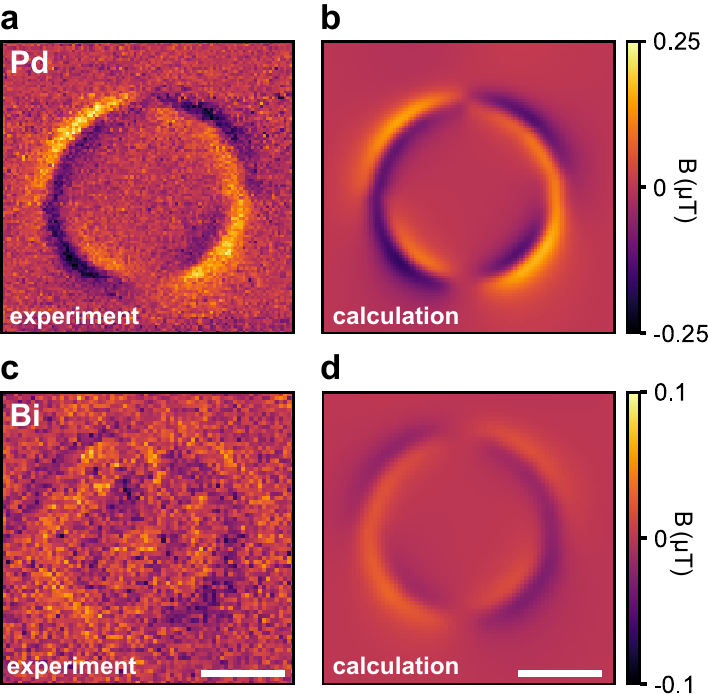}
	\caption{{\bf Nanoscale susceptometry}.
		{\bf a}, Measured $B_1$ image of a 2-$\um$-diameter paramagnetic Pd disc in a $35\unit{mT}$ bias field.
		{\bf b}, Calculated $B_1$ image (best fit) of the Pd disc.
		{\bf c,d}, Corresponding measurement and calculation for diamagnetic Bi in a bias field of $33\unit{mT}$.
		Offsets of $51 \unit{nT}$ and $38 \unit{nT}$ due to the gradient of the external bias magnet are subtracted from {\bf a} and {\bf c} respectively.
		See Methods for calculation and fit details.
		Scale bars, $1\unit{\um}$.
		\label{fig5}
  }
\end{figure}
We conclude our study by demonstrating nanoscale imaging of magnetic susceptibility.  Susceptometry measurements are important for investigating, for example, the magnetic response of patterned metals and materials \cite{gardner01,bluhm09spin},  superconductors \cite{low21} as well as para- and superparamagnetic nanoparticles \cite{gould14,fescenko19}.  Fig.~\ref{fig5}a shows a gradiometry map of a 50-nm-thick disc made from paramagnetic Pd placed in a bias field of $\Bext = 35 \unit{mT}$.  Under the bias field, the disc develops a magnetization of $M = \chi \Bext/\mu_0$, where $\xpd$ is the magnetic susceptibility of the Pd film.
A fit to the data (see Methods), shown in Fig.~\ref{fig5}b, produces a susceptibility of $\xpd = (6.6\pm 0.2)\ee{-4}$.  This is slightly smaller than the value of pure Pd ($\xpd =7.66\ee{-4}$ \cite{crc04}). The decreased susceptibility may be attributed to either a finite-size effect \cite{macfarlane13} or hydrogen adsorption \cite{jamieson72}.  Repeating the same experiment with Bi, a diamagnetic sample, produces an experimental value of $\xbi = -(1.7\pm 0.1)\ee{-4}$, which matches the accepted room temperature value of $\xbi = -1.67\ee{-4}$ \cite{crc04}.  Despite the $\sim 4\times$ weaker susceptibility, the magnetic pattern is clearly visible and its sign inverted compared to the paramagnetic Pd disc.  Additionally, local structure is apparent at the center of the disc that is explained by a variation in the film thickness (Fig.~S13).  Together, Fig.~\ref{fig4}a-d demonstrate the feasibility of extending sensitive dc susceptometry to the nanometer scale. 

\vspace{0.5cm}
\textit{Conclusions -- }
In conclusion, our work demonstrates a simple yet powerful method for the imaging of static magnetization patterns with high sensitivity and spatial resolution.  While we demonstrate scanning gradiometry on a layered antiferromagnet, extension to more challenging systems including perfectly compensated collinear antiferromagnets, screw and step dislocations \cite{ravlic03}, or isolated magnetic defects is natural.  In particular, the magnetic signal from an atomic step edge is equivalent to that of a one-dimensional spin chain with a linear magnetization density of $\sigma_z h \approx 0.5\unit{\muB/nm}$ (Methods), demonstrating the feasibility of imaging generic 1D spin systems.  Gradiometry is also well-positioned for imaging the internal structure of domain walls and skyrmions, and in particular, for quantifying their size and chirality \cite{tetienne15,dovzhenko18,wornle21,velez22} (Supplementary Fig.~S14).
Finally, although we demonstrate gradiometry on magnetic fields, the technique can be extended to electric fields by orienting the external bias magnetic field perpendicular to the NV axis \cite{dolde11,barson21,bian21}.  In particular, the dynamic mode of operation may alleviate charge screening that has previously limited scanning NV electrometry of dc electric field sources \cite{oberg20}.  With the ability to image both magnetic and electric fields, one could imagine correlating antiferromagnetic and ferroelectric order in multiferroics \cite{cheong20}, providing a unique angle to investigate the magneto-electric coupling in these fascinating and technologically important materials.


\vspace{0.5cm}
{\bf Acknowledgments}

The authors thank P. Scheidegger, S. Diesch, S. Ernst, K. Herb and M. S. W\"ornle for fruitful discussions, 
J. Rhensius for help with the Pd and Bi micro-disc fabrication and for the scanning electron micrograph in Fig.~\ref{fig1}c,
M. Giraldo and M. Fiebig for providing the \CrO\ samples, 
P. Gambardella for helpful comments on the manuscript, 
and the staff of the FIRST lab cleanroom facility for technical support.
This work was supported by the European Research Council through ERC CoG 817720 (IMAGINE), Swiss National Science Foundation (SNSF) Project Grant No. 200020\_175600, the National Center of Competence in Research in Quantum Science and Technology (NCCR QSIT), and the Advancing Science and TEchnology thRough dIamond Quantum Sensing (ASTERIQS) program, Grant No. 820394, of the European Commission.
M. T. acknowledges the Swiss National Science Foundation under Project No. 200021-188414.

\vspace{0.5cm}
{\bf Author contributions}

C.L.D., W.S.H. and M.L.P. conceived the idea and developed the theory.
W.S.H., M.L.D. and M.L.P. implemented the scanning gradiometry technique, carried out the magnetometry experiments and performed the data analysis.
P.W. provided technical support with the magnetometry experiments.
M.T. performed the atomic force microscopy of the unpolished \CrO\ crystal.
M.L.P. prepared the bilayer graphene sample.
C.-H. L. prepared the Pd and Bi micro-discs.
C.L.D. supervised the work.
W.S.H. and C.L.D. wrote the manuscript. All authors discussed the results.

\vspace{0.5cm}
{\bf Competing interests}

The authors declare no competing interests.

\vspace{0.5cm}
{\bf Additional information}

Supplementary information accompanies this paper.
Correspondence and requests for materials should be addressed to C.L.D.

	
\clearpage
{\bf Methods}
	
\vspace{0.5cm}
{\bf Experimental set-up}

All experiments were carried out at room temperature with two home-built scanning microscopes. Micro-positioning was carried out by closed-loop three-axis piezo stages (Physik Instrumente) and AFM feedback control was carried out by a lock-in amplifier (HF2LI, Zurich Instruments) and standard PID controls. PL of the NV centers was measured with avalanche photodiodes (APDs) (Excelitas) and data were collected by standard data acquisition cards (PCIe-6353, National Instruments). Direct currents sent through the bilayer graphene were created with an arbitrary waveform generator (DN2.663-04, Spectrum Instrumentation) and the current was measured with a transimpedance amplifier (HF2TA, Zurich Instruments). Microwave pulses and sequences were created with a signal generator (Quicksyn FSW-0020, National Instruments) and modulated with an IQ mixer (Marki) and an arbitrary waveform generator (DN2.663-04, Spectrum Instrumentation and HDAWG, Zurich Instruments). NV centers were illuminated at $<100\unit{\uW}$ by a custom-designed $520\unit{nm}$ pulsed diode laser. Scanning NV tips were purchased from QZabre AG \cite{qzabre}. Three different NV tips were used throughout this study with stand-off distances $d$ between $70-130\unit{nm}$ (excluding the 20 nm retract distance).

{\bf Magnetic samples}

\noindent\textit{Bilayer graphene device:} Standard microfabrication processes were used, including mechanical exfoliation and a dry transfer process for generating the hBN-bilayer-graphene-hBN stack, electron beam lithography and plasma etching for creating the device geometry and physical vapor deposition for creating the metallic device contacts. Full details can be found in Ref.~\onlinecite{palm22}.
	
\noindent\textit{\CrO\ crystals:} The mechanically polished and non-polished \CrO\ bulk single crystals were provided by Prof. Manfred Fiebig. Both crystals have a (0001) surface orientation.  Crystal growth and processing details can be found in Refs.~\onlinecite{wornle21,fiebig96thesis}. 
	
\noindent\textit{Pd and Bi discs:}  Metallic micro-discs were created using standard microfabrication processes. Discs were defined through electron beam lithography on spin-coated Si wafers. Chemical development of the spin-coated resist, followed by metal deposition of either 50 nm of Pd or Bi and a lift-off process finished the fabrication.

{\bf Initialization and readout of NV spin state}

A laser pulse of $\sim 2\unit{\us}$ duration was used to polarize the NV center into the $m_S=0$ state. Then, the dc or ac quantum sensing measurement occurred on the $m_S=0$ to $m_S=-1$ transition (see below). The readout of the NV's spin state was performed by another $\sim 2\unit{\us}$ long laser pulse, during which the photons emitted from the NV center were collected with the APD, binned as a function of time, and summed over a window ($\sim 300 \unit{ns}$) that optimized spin-state-dependent PL contrast \cite{schirhagl14}.
	
{\bf DC sensing protocol}

DC magnetic images were acquired with the pulsed ODMR method \cite{dreau11}.  Obtained magnetic resonance spectra were fitted for the center frequency $f_c$ of a Lorentzian function,
$L(f) = 1 - \epsilon [(2\pi f -2\pi f_c)/\omega_L]^2+1]^{-1}$ ,
where $\epsilon$ is the spin contrast (in percent) and $\omega_L$ is the width of the Lorentzian dip. In a 2D scan the magnetic field projected along the axis of the NV could be determined at each pixel using
$B(x,y) = 2 \pi [f_c (x, y) - f_0]/\ye$
where $f_0$ is the resonance frequency far from the surface.
For our diamond probes, the NV anisotropy axis was at an $\theta \sim 55^\circ$ angle with respect to the out-of-plane direction ($z$-axis in Fig.~1).  Therefore, $B(x,y)$ corresponded to the vector field projection along this tilted direction.
	
{\bf AC sensing protocol}

AC sensing used either a spin echo and or dynamical decoupling sequence \cite{taylor08,delange11}.  The NV spin state was initialized optically into the $m_S=0$ state followed by a microwave $\pi/2$-pulse to create a coherence between the $m_S = 0$ and $m_S =-1$ states.  The quantum phase $\phi$ accumulated between the two states during the coherent precession can be expressed as
$\phi = \int_0^{\tau} \ye g(t) B(t) dt$
where $g(t)$ is the modulation function \cite{degen17}, $B(t)$ is the magnetic field, $\tau$ is the interaction time (the time in between the first and last $\pi/2$-pulse), and where we use the rotating frame approximation.  The modulation function alternates between $\pm1$ with each microwave $\pi$-pulse during the pulse sequence. While imaging with the gradiometry pulse sequences, $3\pi/2$-pulses were sometimes used instead of the final $\pi/2$-pulses for the projective spin readout to reduce pulse imperfections caused by (ca. $\pm 100\unit{kHz}$) drifts in the NV resonance frequency. 
	
{\bf Calibration of tuning fork oscillation}

We estimated the oscillation amplitude $\xosc$ and oscillation angle in the $xy$-plane, denoted by $\alpha$, of the tuning forks with two different \textit{in-situ} measurement techniques.   The first method involved processing a static field map and a gradient field map acquired over the same region with a least-squares minimization scheme. By minimizing a cost function proportional to the pixel differences between the $B_1$ and numerically differentiated $B_0$ images, estimates for $\xosc$ and $\alpha$ were determined. The second method involved a stroboscopic imaging of the static field ($B_0$) synchronized to the tuning fork oscillation. By measuring the time-tagged displacements of magnetic features recorded at different positions during the tuning fork oscillation, $\xosc$ and $\alpha$ could be estimated by fitting to the path taken by the NV. We found that $\xosc$ scaled linearly with the applied drive voltage for the voltages we used while imaging.  Examples of these calibration methods are given in Supplementary Note 4.

{\bf Calibration of the trigger delay}

In order to optimize the ac sensing sequences, and to distinguish between signals from the first and second derivatives (see Supplementary Note 1), a trigger delay ($t_0$) calibration measurement must be made.  The calibration measurement consisted of measuring the phase $\phi$ at a stationary point on the sample, while varying $t_0$, thus producing a phase that oscillated as a function of $t_0$. The chosen value of $t_0$ was the value which maximized the measured phase. Examples of this calibration measurement, as well as the characterization of PL oscillation caused by the tuning fork motion, are shown in the Supplementary Note 4.
	
{\bf Reconstruction of the static field map from the gradient map}

A gradient map $B_1$ can be used to reduce the noise and improve the image of a (less sensitive) static field map $B_0$. Letting $\BT$ denote the true static magnetic field value, the experimentally measured $B_0$ and $B_1$ field maps are 
\begin{equation}
	\begin{split}
		B_0 (x, y) &= \BT(x, y) + w_0\\
		B_1 (x, y) &= \xosc \frac{\partial}{\partial r} \BT(x, y) + w_1
	\end{split}
\end{equation}
where $\frac{\partial}{\partial r} = \cos(\alpha)\frac{\partial}{\partial x} + \sin(\alpha)\frac{\partial}{\partial y}$ is the directional derivative and $w_0 \sim \mathcal{N}(0, \sigma_{B_0}^2) $ and  $w_1 \sim \mathcal{N}(0,\sigma_{B_1}^2) $ are white noise added to each pixel (reflecting the Poissonian shot noise of the photo-detection). In $k$-space these equations can be transformed into
\begin{equation}
	\begin{split}\label{eq:Bfft}
		\hat{B}_0 (\kx, \ky) &= \hatBT(\kx, \ky) +\hat{w}_0\\
		\frac{\hat{B}_1 (\kx, \ky)}{i\xosc \kr}  &= \hatBT(\kx, \ky) + \frac{\hat{w}_1}{i\xosc \kr}
	\end{split}
\end{equation}
where $\hat X$ denotes the Fourier transform of $X$, $\kr = \kx \cos\alpha + \ky \sin\alpha$ is the dot product between the oscillation direction and the $k$-vector $[\kx, \ky]$. Integration in $k$-space introduces a $k$-dependent noise term in the gradient field map. In particular, the integrated gradient map has noise amplification near the line $\ky = -\cot(\alpha)\kx$ but has noise suppression far away from that line. Directional sensitivity could be avoided by oscillating the tuning fork in the $z$-direction (tapping mode), or by using an oscillator that supports orthogonal lateral modes. To circumvent this problem we average two $k$-space maps (Eq.~\ref{eq:Bfft}) with $k$-vector dependent weights that reflect the noise added by the integration process. The use of inverse variance weights additionally results in an image with the lowest possible variance. Thus, the optimal reconstructed static field map can be computed by taking the inverse Fourier transform of
\begin{align}
\hat{\tilde B}_0(\kx,\ky) = \hat B_0(\kx,\ky)\, \frac{\ko^2}{\ko^2+\kr^2} + \frac{\hat B_1(\kx,\ky)}{i\xosc\kr}\, \frac{\kr^2}{\ko^2 +\kr^2}
\end{align}
where $\ko^2 = \sigma_{B_1}^2/(\xosc^2\sigma_{B_0}^2)$ defines a cut-off wave vector determined by the oscillation amplitude and the noise variances $\sigma_{B_0}^2$ and $\sigma_{B_1}^2$.
The corresponding cut-off wavelength $\lambda = 2\pi/\ko$ reflects the spatial wavelength above (below) which the $B_0$ ($B_1$) field map is less noisy. It should also be noted that the $k$-space averaging process can be modified to include the $B_2$ map, however the $B_1$ map provides the most significant improvement. 
Specifically, in Fig.~\ref{fig4}c, $\xosc = 23\unit{nm}$ and $\alpha = 180^\circ$ were used as reconstruction parameters (determined by the calibration in Supplementary Note 4). Note, since the gradient and $k$-space averaging are directional, the noise reduction is also directional (approximately $22\times$ in the $x$-direction and $5.6\times$ in the $y$-direction).

{\bf Stray fields from atomic step edges}

For an atomic step edge propagating along the $y$-direction the stray field is modeled by two out-of-plane magnetic samples with different heights. Taking the analytical form (see Ref.~\onlinecite{tetienne15}) for the stray fields above an edge at $x = 0$ as $B_x (x, z) = \frac{-\mu_0 \sigma_z}{2 \pi}\frac{z}{x^2 + z^2}$,  $B_y = 0$, and $B_z (x, z) = \frac{\mu_0 \sigma_z}{2 \pi}\frac{x}{x^2 + z^2}$, where $\sigma_z$ is the surface magnetization, we define the stray field produced by a small change in the height $h$ as 
\begin{equation}
	\begin{split}
		B_x^\mr{step} &= B_x (x, d) - B_x (x, d+h) \approx \frac{\mu_0 \sigma_z h}{2 \pi}\frac{(x^2 - d^2)}{(x^2 + d^2)^2} \\
		B_z^\mr{step} &= B_z(x, d) - B_z (x, d+h) \approx \frac{\mu_0 \sigma_z h}{2 \pi}\frac{2xd}{(x^2 + d^2)^2}
	\end{split}\label{eq:steps}
\end{equation}
with $B_y^\mr{step} = 0$. The expressions are simplified in the limit of $h \ll d$ since the sub-nanometer atomic step edges are much smaller than typical standoff distances of $d \sim 50-100 \unit{nm}$. We measure the gradient of the stray field along the oscillation direction, taken as the $x$-direction for simplicity. This leads to field gradients of
\begin{equation}
	\begin{split}
		\frac{\partial B_x^\mr{step}}{\partial x} &= \frac{\mu_0 \sigma_z h}{\pi}\frac{x(3d^2 - x^2)}{(x^2 + d^2)^3} \\
		\frac{\partial B_z^\mr{step}}{\partial x} &= \frac{\mu_0 \sigma_z h}{\pi}\frac{d(d^2 - 3x^2)}{(x^2 + d^2)^3}
	\end{split}
\end{equation}
The fits in Fig.~\ref{fig4}g and ~\ref{fig4}h are produced by projecting the field gradients onto the NV axis defined by $\vec{e} = [\sin\theta\cos\varphi, \sin\theta\sin\varphi, \cos\theta]$ and summing over multiple steps. The line cuts are produced by rotating and plane averaging a 2D image (Fig.~S12). We account for the rotation angle in the $xy$-plane by introducing an additional image rotation angle $\varphi'$. The experimentally measured $B_1$ field is then fitted by
\begin{align}
	\frac{B_1}{\xosc\cos(\varphi')} = \sum_\mr{steps}  \sin(\theta)\cos(\varphi+\varphi')\frac{\partial B_x^\mr{step}}{\partial x}  +  \cos(\theta) \frac{\partial B_z^\mr{step}}{\partial x}
\end{align}
We set $h=0.227\unit{nm}$ and $\varphi' = 40^\circ$ for the fit in Fig.~\ref{fig4}g, $h=0.454\unit{nm}$ and  $\varphi' = 23^\circ$ for the fit in Fig.~\ref{fig4}h and fixed $\xosc = 46\unit{nm}$ (determined by the tip calibration) in both fits. The standoff distance $d$, the surface  magnetization $\sigma_z$, angles $\theta$ and $\varphi$ and step edge locations are free parameters in the fit. Collectively, the fitted line scans give $d = 89 \pm 12 \unit{nm}$ (including the $20 \unit{nm}$ retract distance) and $\sigma_z = 2.1 \pm 0.5 \unit{\mu_B/nm^2}$, which is consistent with previous measurements on this sample \cite{wornle21}.

Note, Eq.~\ref{eq:steps} is functionally equivalent to the magnetic field from a one-dimensional ferromagnetic spin chain with a linear magnetization density of $M^{1D} = \sigma_z h$. To demonstrate this, we calculate the stray field produced by a infinite line of magnetic dipoles along the line $x = 0$ and pointing in the $x$-direction as:
\begin{equation}
	\begin{split}
		B_x^\mr{1D} &= \frac{\mu_0 M^\mr{1D}}{4\pi}\int_{-\infty}^{\infty} \left(\frac{3x^2}{r^5} - \frac{1}{r^3} \right) dy = \frac{\mu_0 M^\mr{1D}}{2\pi}\frac{(x^2 - d^2)}{(x^2 + d^2)^2} \\
		B_z^\mr{1D} &= \frac{\mu_0 M^\mr{1D}}{4\pi}\int_{-\infty}^{\infty} \frac{3xd}{r^5} dy = \frac{\mu_0 M^\mr{1D}}{2\pi}\frac{2xd}{(x^2 + d^2)^2}
	\end{split}
\end{equation}
where $r^2 = x^2 + y^2 + d^2$. 

{\bf Susceptibility fitting for Pd and Bi discs}

Susceptibility fits followed the assumption that the stray field produced by a para- or diamagnetic sample is identical to that of a homogeneously magnetized body with a magnetization magnitude of $M = \chi_\mr{Pd/Bi} B_\mr{ext}/\mu_0$ and a magnetization vector that is parallel to the external polarizing field.
The stray field produced by this magnetization is $\vec{B}(\vec{r}) = -\mu_0 \nabla \phi_\mr{mag}(\vec{r})$ where $\phi_\mr{mag}(\vec{r})$ is the magnetic potential.
We computed the stray field using the $k$-space method of Ref.~\cite{beardsley89} and fitted the gradient image to the numerically computed gradient fields. In the fitting procedure we fixed $\xosc= 69\unit{nm}$, $\alpha = 180^\circ$ (determined by the tip calibration) and sample thickness $t = 50\unit{nm}$ while the standoff distance $d$, magnetization magnitude $M$, angles $\theta$ and $\varphi$, circle radius and position are free parameters in the fit. The susceptibilities are computed as $\chi_\mr{Pd/Bi} = \mu_0 M / B_\mr{ext}$ and the fitted magnetizations were $M_\mr{Pd} = 18.6 \pm 0.6 \unit{A/m}$ and $M_\mr{Bi} = -4.4 \pm 0.4 \unit{A/m}$. 


%

\end{document}